\documentclass{aa}
\usepackage{natbib}

\usepackage{bm}
\usepackage{amsmath}

\usepackage{color}
\usepackage{ulem}

\begin{document}

%
   \title{Double plasma resonance instability as a source of solar zebra emission}

   \authorrunning{Ben\'a\v{c}ek and Karlick\'y}
   \titlerunning{Double plasma resonance instability}

   \author{J. Ben\'a\v{c}ek$^{1}$ and M. Karlick\'y$^2$
          }
   \offprints{, \email{jbenacek@physics.muni.cz}}

   \institute{Department of Theoretical Physics and Astrophysics, Masaryk University,
   Kotl\'a\v{r}sk\'a 2, CZ -- 611 37 Brno, Czech Republic
         \and
   Astronomical Institute of the Czech Academy of Sciences, Fri\v{c}ova 258, CZ -- 251 65 Ond\v rejov, Czech Republic\\}
   \date{Received ; accepted }


  \abstract
   {The double plasma resonance (DPR) instability plays a basic role in the generation of solar radio zebras.
   In the plasma, consisting of the loss-cone type distribution of hot electrons and much denser and colder
   background plasma, this instability generates the upper-hybrid waves, which
   are then transformed into the electromagnetic waves and observed as radio zebras.}
  {In the present paper we numerically study the double plasma resonance instability from the point of view
  of the zebra interpretation.}
   {We use a 3-dimensional electromagnetic particle-in-cell (3-D PIC) relativistic model.
   We use this model in two versions: a) a spatially extended "multi-mode" model and
   b) a spatially limited "specific-mode" model.
   While the multi-mode model is
   used for detailed computations and verifications of the results obtained by
   the "specific-mode" model, the specific-mode model is used for computations
   in a broad range of model parameters, which considerably save computational time.
   For an analysis of the computational results, we developed software tools in Python.}
   {First using the multi-mode model, we study details of the double plasma resonance instability.
   We show how the distribution function of hot electrons changes during this
   instability.
   Then we show that there is a very good agreement between results obtained by
   the multi-mode and specific-mode models, which is caused by a dominance
   of the wave with the maximal growth rate. Therefore, for computations in a
   broad range of model parameters, we use the specific-mode model.
   We compute the maximal growth rates of the double plasma resonance instability
   with a dependence on the ratio between the upper-hybrid $\omega_\mathrm{UH}$
   and electron-cyclotron $\omega_\mathrm{ce}$ frequency.
   We vary temperatures of both the hot and background plasma
   components and study their effects on the resulting growth rates.
   The results are compared with the analytical ones.
   We find a very good agreement between numerical and analytical growth rates.
   We also compute saturation energies of the upper-hybrid waves in a very broad range of
   parameters.
   We find that the saturation energies of the upper-hybrid waves show maxima and minima
   at almost the same values of $\omega_\mathrm{UH}/\omega_\mathrm{ce}$ as the growth rates,
   but with a higher contrast between them than the growth rate maxima and
   minima. The contrast between saturation energy maxima and minima increases when the temperature of hot electrons increases.
   Furthermore, we find that the saturation energy of the upper-hybrid waves is proportional
   to the density of hot electrons. The maximum saturated energy can be up to one percent of the kinetic
   energy of hot electrons. Finally we find that the saturation energy maxima in the interval of $\omega_\mathrm{UH}/\omega_\mathrm{ce}$ = 3-18
   decrease according to the exponential function. All these findings can be used in
   the interpretation of solar radio zebras.}
   {}

   \keywords{Instabilities -- Methods: numerical -- Sun: radio radiation}

   \maketitle

\section{Introduction}

The loss-cone type of distribution of hot electrons superimposed on the
distribution of much denser and colder background plasma is unstable due to the
double plasma resonance instability
\citep{1975SoPh...44..461Z,1982ApJ...259..844M,
1983SoPh...88..297Z,1986ApJ...307..808W,2001SoPh..202...71L,
2013SoPh..284..579Z,2015A&A...581A.115K}. This instability generates the
upper-hybrid waves, which have their maxima close to the gyro-harmonic number $s$ =
$\omega_\mathrm{UH}/\omega_\mathrm{ce}$, where $\omega_\mathrm{UH}^2 =
\omega_\mathrm{pe}^2 + \omega_\mathrm{ce}^2$ and  $\omega_\mathrm{UH}$,
$\omega_\mathrm{pe}$ , and $\omega_\mathrm{ce}$ are the upper-hybrid,
electron-plasma, and electron-cyclotron frequency, respectively. Owing to this
property, the double plasma resonance (DPR) instability is used in models of solar
radio zebras
\citep{1975SoPh...44..461Z,1986ApJ...307..808W,2001SoPh..202...71L,
2011AnGeo..29.1673T,2013SoPh..284..579Z,2015A&A...581A.115K,2015SoPh..290.2001Y,2017A&A...555A...1B}.

The resonance condition for the double plasma resonance instability in the relativistic case can be expressed as
\begin{equation}
\omega_\mathrm{UH} - \frac{k_\parallel u_\parallel}{\gamma} - \frac{s \omega_\mathrm{ce}}{\gamma} = 0,
\end{equation}
where $\mathbf{k} = (k_\parallel, k_\perp)$ is wave vector,
$\mathbf{u} = (u_\perp, u_\parallel)$, $u_\perp = p_\perp/m_\mathrm{e}$ , and
$u_\parallel = p_\parallel/m_\mathrm{e}$ are the hot electron velocities
perpendicular and parallel to the magnetic field; $m_e$ is the electron mass,
$\gamma = \sqrt{1 + \mathbf{u}^2/c^2}$ is the relativistic Lorentz factor, $s$
is the gyro-harmonic number, and $c$ is the speed of light. In theoretical
models studying the double plasma resonance instability, the distribution of hot
electrons is usually taken as the Dory-Guest-Harris (DGH, \citet{1965PhRvL..14..131D})
distribution  for $j=1$ in the form
\begin{equation}
f = \frac{u_\perp^2}{2 (2\pi)^{3/2} v_\mathrm{t}^5} \exp \left(-\frac{u_\perp^2 + u_\parallel^2}{2 v_\mathrm{t}^2}\right),
\label{edory}
\end{equation}
where $v_\mathrm{t}$ we call the thermal velocity of hot electrons, although
the distribution function in the relation~\ref{edory} is not Maxwellian. The
distribution of the background plasma is taken as a Maxwellian one.
The combination of both these distributions is considered to be the
prototype distribution generating radio zebras \citep{1986ApJ...307..808W}.

In our previous paper~\citep{2017A&A...555A...1B} we studied the double plasma resonance instability analytically. We showed that the maxima of growth rates of the upper-hybrid waves
are shifted to lower ratios of $\omega_\mathrm{UH}/\omega_\mathrm{ce}$ and the contrast between maxima and minima of the growth rate decreases as the temperature of hot electrons increases.
On the other hand, when the temperature of the background plasma is increased, the contrast remains the same.

In studies of solar radio zebras \citep[e.g.][]{2001SoPh..202...71L},
the frequencies of the zebra stripes and the contrast of these stripes to the
background continuum are analyzed. The frequencies are used for the determination
of the magnetic field in zebra radio sources. Therefore, it is important to
know the relation between the zebra stripe frequencies and the gyro-harmonic
number $s \approx \omega_\mathrm{UH}/\omega_\mathrm{ce}$. Furthermore, the contrast
of the zebra stripes is believed to be connected with the temperature of hot
electrons \citep{2004SoPh..219..289Y}. Moreover, some zebras have many stripes
and their frequencies correspond to high gyro-harmonic numbers (sometimes $>$
20) \citep{2015A&A...581A.115K}. All these facts require computations in a very
broad range of the gyro-harmonic number $s$, which are difficult to make with
the spatially extended multi-mode model. Fortunately,  we found that the maximal
growth rates and saturation energies computed in the multi-mode model agree
very well with that computed in the spatially limited specific-mode model. This
agreement is caused by the fact that the wave with the maximal growth rate
becomes very soon dominant over other growing waves (due to its exponential
growth). Using this specific-mode model, which considerably saves computational
time, we compute the maximal growth rates and saturation energies of the DPR
instability in a broad range of parameters. Because in the analytical analysis
of growth rates of the DPR instability several assumptions were made, these
numerical computations serve to verify the analytical results.

The paper is structured as follows. In Section 2 we describe our numerical
model and initial conditions for the studied double plasma resonance
instability. In Section 3 we present the results for different model
parameters.  Discussion of these results and conclusions are in Section 4.

\section{Model}
We use a 3-D particle-in-cell (PIC) relativistic model
\citep{1985stan.reptR....B,matsumoto.1993,2008SoPh..247..335K} with multi-core Message Passing Interface (MPI) parallelization. Further details can be found in \citet[p.67-84]{matsumoto.1993} and
on the link below.\footnote{https://www.terrapub.co.jp/e-library/cspp/text/10.txt.} In
the present article we use this model in two versions: a) the spatially extended
model with many wave modes (multi-mode model), and b) spatially limited model
with specific wave mode (specific-mode model), which is used for relatively
fast computing in the broad range of model parameters. The model size
 in x-, y-, and z-directions is $128 \times 60 \Delta \times 128
\Delta$ for the multi-mode model, and   $\lambda \Delta \times \lambda \Delta
\times 32 \Delta$ for the specific-mode model, respectively, where $\Delta=1$
is the grid size and $\lambda$ is the wavelength of the specific upper-hybrid
wave. For chosen gyro-harmonic numbers and plasma temperatures, we fit the size
of the specific-mode model to find the wave mode with the maximal growth rate.
In all models we use periodic boundary conditions in all three directions.

The model time step is $\mathrm{d}t = 1$, electron plasma frequency
$\omega_\mathrm{pe} = 0.05$, initial magnetic field is in the $z-$ direction, and
electron cyclotron frequency $\omega_\mathrm{ce}$ varies from
$0.38~\omega_\mathrm{pe}$ to $0.056~\omega_\mathrm{pe}$ approximating the
gyro-harmonic numbers $s=3-18$. For dependencies of the growth rate on
temperatures, we use $s$ in the range of $s=3-7$. Higher values of $s$ up to 18
are used for calculating the saturation energies of the generated
upper-hybrid waves, which is needed for zebras with many zebra stripes. In the
model we use two groups of electrons: cold background Maxwellian electrons with
the thermal velocity $v_{\mathrm{tb}} = 0.03-0.05$~c, corresponding to the
temperature in the interval 5.4-14.8~MK, and hot electrons with the DGH
distribution having the velocity $v_{\mathrm{t}} = 0.15-0.3$~c. Higher values
of the background plasma temperatures are given by the requirements of the PIC
model. However, as known from the previous analytical
study~\citep{2017A&A...555A...1B} and shown in the following, variations of the
background plasma temperature have little effect.

The electron density of background electrons was taken as $n_\mathrm{e} = 960$
per cell and the ratio of background to hot electrons was taken as
$n_\mathrm{e}/n_\mathrm{h} = 8$ with some exceptions mentioned in the
following. The number of protons is the same as the number of electrons and their
temperature is always the same as that of background electrons. The
proton-to-electron mass ratio $m_\mathrm{p}/m_\mathrm{e}$ is 1836.

To find the maximal growth rate for specific physical parameters, we changed the
model size $\lambda \Delta$ in the perpendicular direction to the magnetic
field. Namely, we fit the wavelength of the upper-hybrid wave with the maximal
growth rate to a distance of grids used in the PIC model. Thus, we get the
maximal simulated growth rate. We found that the optimal model size does not
change when computing with different ratios of the electron plasma and electron
cyclotron frequencies, but is changed by changing the temperature.

At the beginning of DPR instability, the electric wave energy grows
exponentially. We fit this part of the electric energy evolution by the
function $W_\mathrm{UH} = A_0 \exp (2 \Gamma t)$,  where $A_0$ is the initial
electric energy, $t$ is time, and $\Gamma$ is the growth rate in model units.
Finally we expressed the growth rate of the upper-hybrid waves in the ratio to
the upper-hybrid frequency.

\section{Results}

\begin{figure}[ht]
\centering
\includegraphics[width=0.45\textwidth]{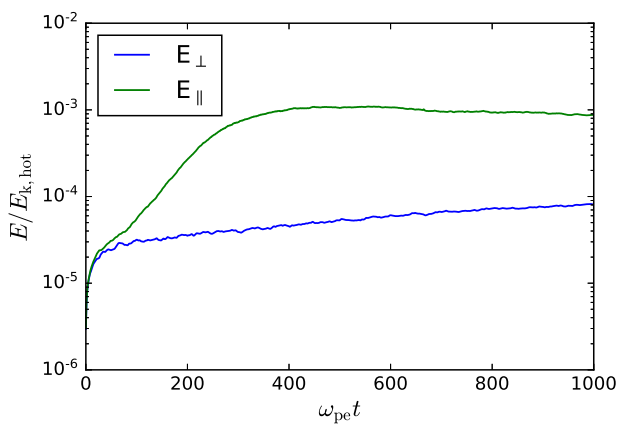}
\caption{Time evolution of the electric energy density in the parallel and perpendicular
directions to the magnetic field lines in Model~2M (Table~1),
normalized to the kinetic energy density of hot electrons for the maximal growth rate
and the gyro-harmonic number $s=6$.}
\label{fig1a}
\end{figure}

Firstly we compute an evolution of the DPR instability and generation
of the upper-hybrid waves in the multi-mode models with the parameters
summarized in Table~1. We show an example of the time evolution of the electric energy
density in the parallel and perpendicular directions to the magnetic field
lines in Model~2M normalized to the kinetic energy density of hot electrons for
the maximal growth rate and the gyro-harmonic number $s=6$  in
Figure~\ref{fig1a}. As can be seen here, the electric energy density in the
perpendicular direction dominates over the parallel energy density by at least
two orders of  magnitude, which indicates the generation of the upper-hybrid waves.

\begin{table}[!ht]
\caption{Parameters of the multi-mode models.} \label{tab1} \centering
\begin{tabular}{cccr}
\hline\hline
Comp. no. &  $v_\mathrm{t}$ & $v_\mathrm{tb}$ & $s$ \\
\hline
1M & 0.15~c & $0.03$~c & 6 \\
2M & 0.2~c & $0.03$~c & 6 \\
3M & 0.2~c & $0.03$~c & 3 \\
4M & 0.3~c & $0.03$~c  & 5 \\
\hline
\end{tabular}
\end{table}

\begin{figure}[ht]
\centering
\includegraphics[width=0.45\textwidth]{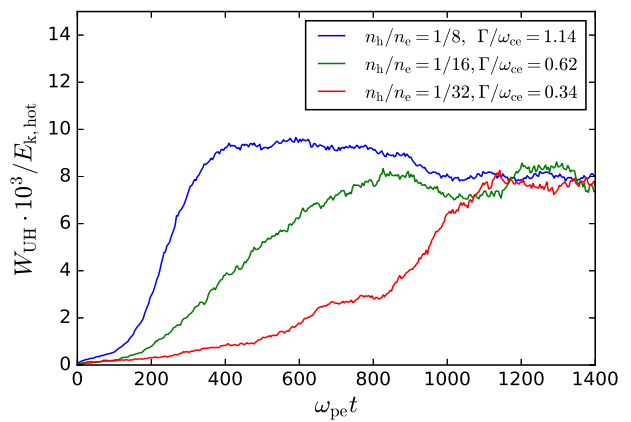}
\caption{Time evolution of the energy density of the upper-hybrid waves normalized
to the kinetic energy density of hot electrons for three values of $n_\mathrm{h}/n_\mathrm{e}$
and for the maximal growth rate and $s=6$. Other parameters are the same as in Model~2M.}
\label{fig1b}
\end{figure}

Examples of the time evolution of the energy density of the upper-hybrid waves
normalized to the kinetic energy density of hot electrons for three
values of $n_\mathrm{h}/n_\mathrm{e}$,  the gyro-harmonic number $s=6$, the
maximal growth rate, and Model~2M are presented in Figure~\ref{fig1b}. The
growth rates, estimated in the early stages of the evolution, are written in this
figure. As seen here, these growth rates are proportional to the density of hot
electrons, in agreement with the paper by \citet{2004SoPh..219..289Y} .
Furthermore, the normalized energy density of the upper-hybrid waves in all
three models converges to the same saturation energy, which indicates that the
saturation energy of the upper-hybrid waves is also proportional to the density
of hot electrons. The kinetic energy density of hot electrons
$E_\mathrm{k,hot}$ depends on the plasma density of hot electrons.

\subsection{DPR instability in detail}

\begin{figure*}[!htb]
\centering
\includegraphics[width=0.49\textwidth]{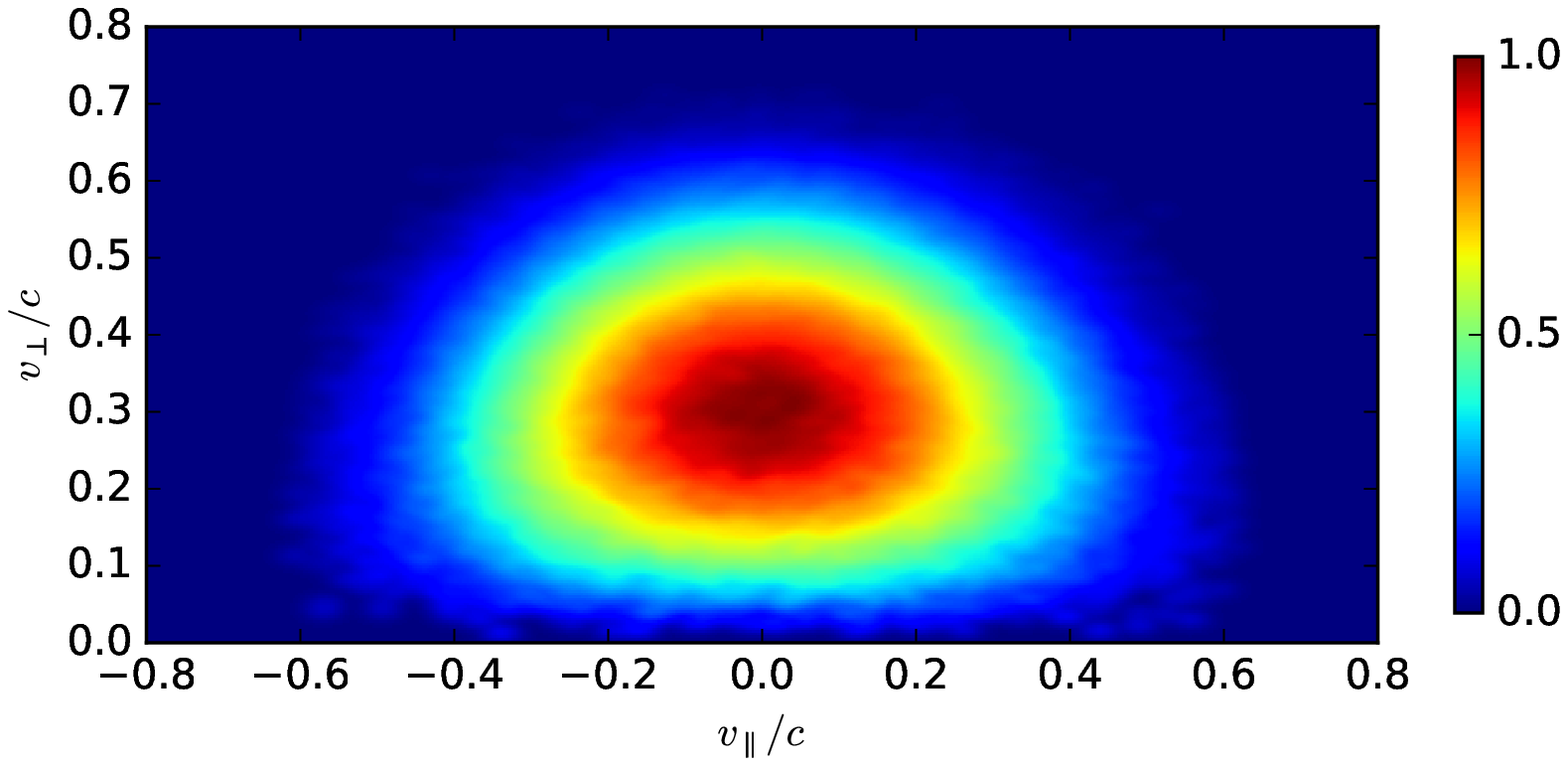}
\includegraphics[width=0.49\textwidth]{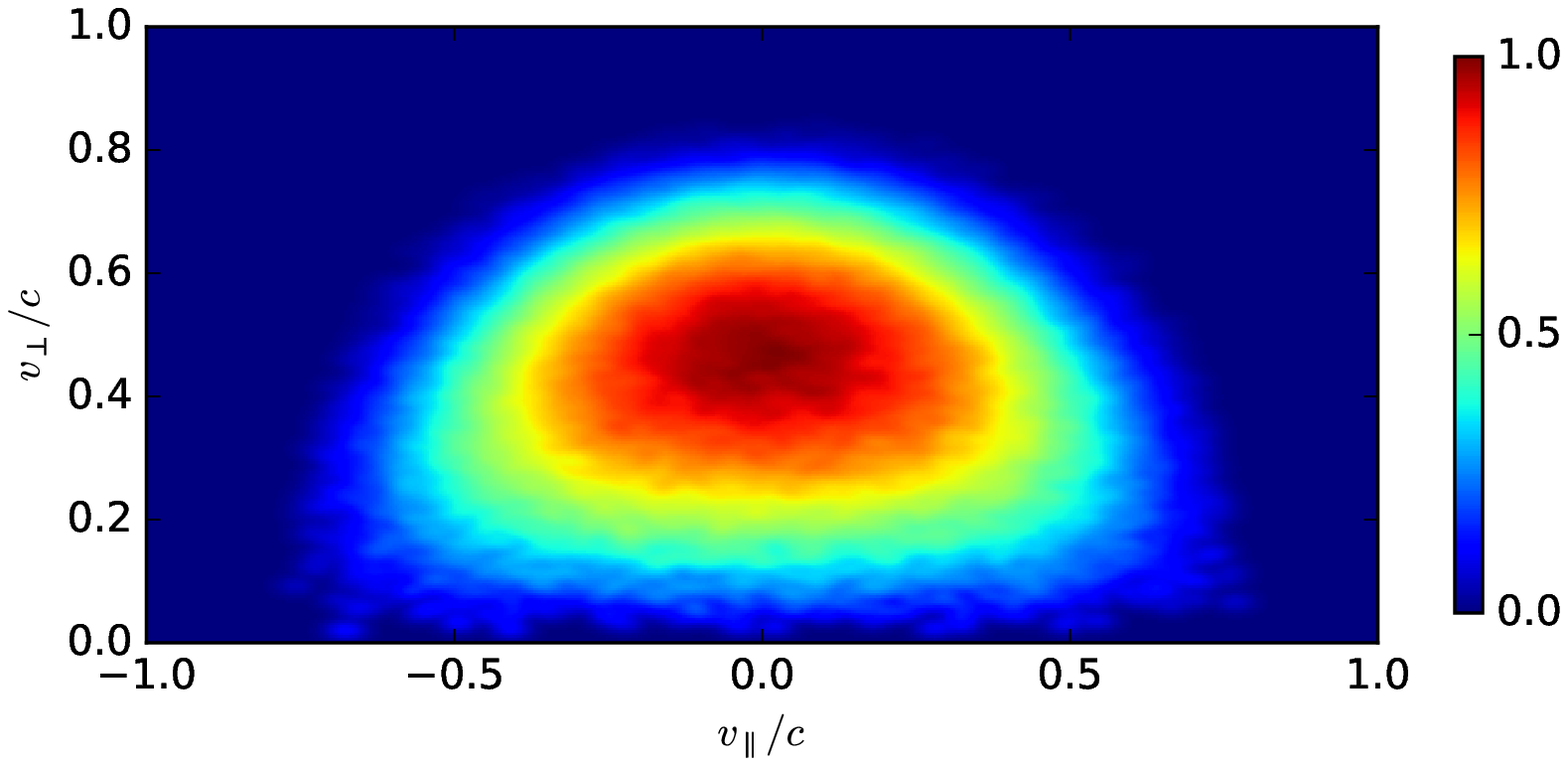}\\
\includegraphics[width=0.49\textwidth]{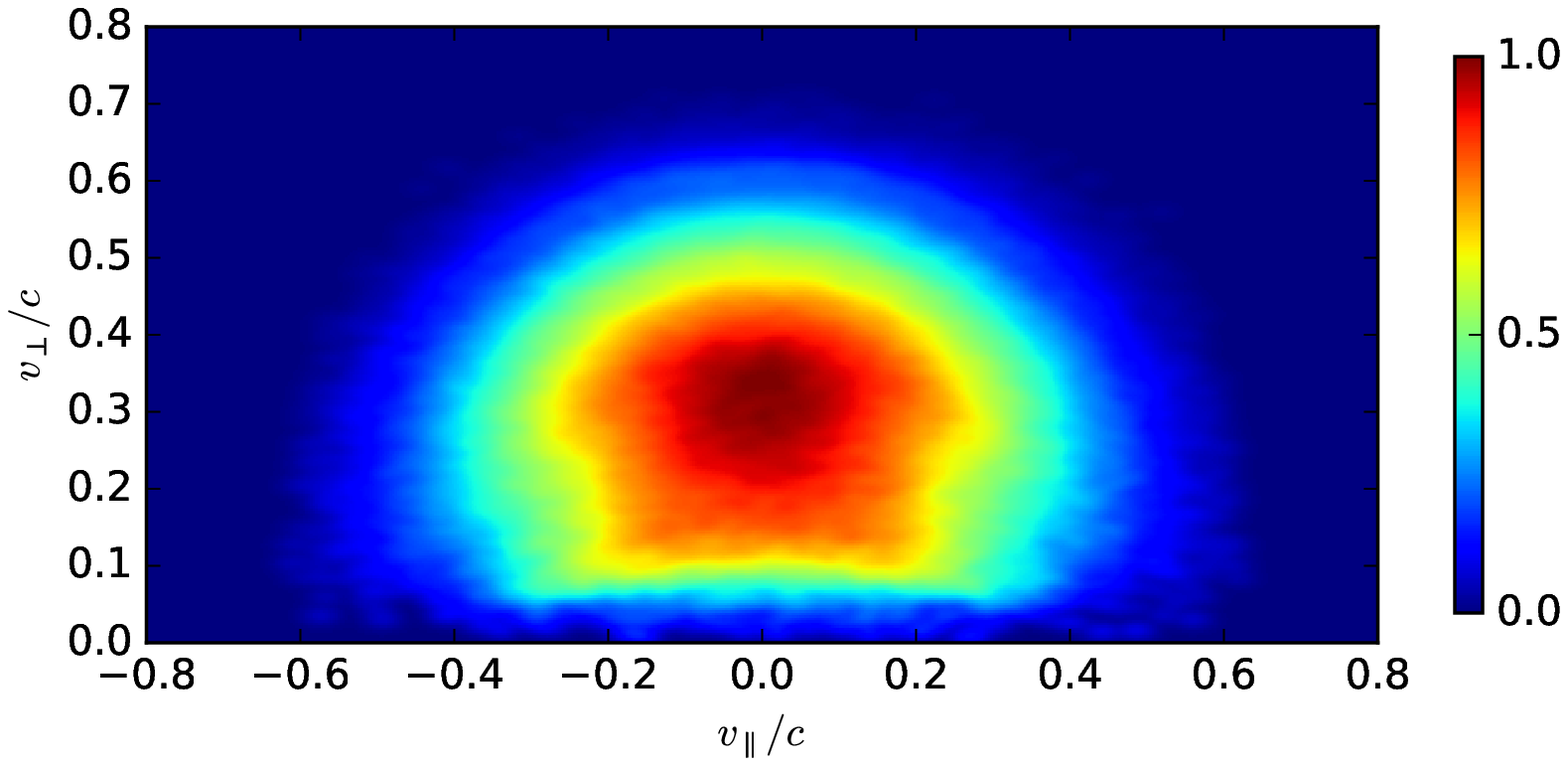}
\includegraphics[width=0.49\textwidth]{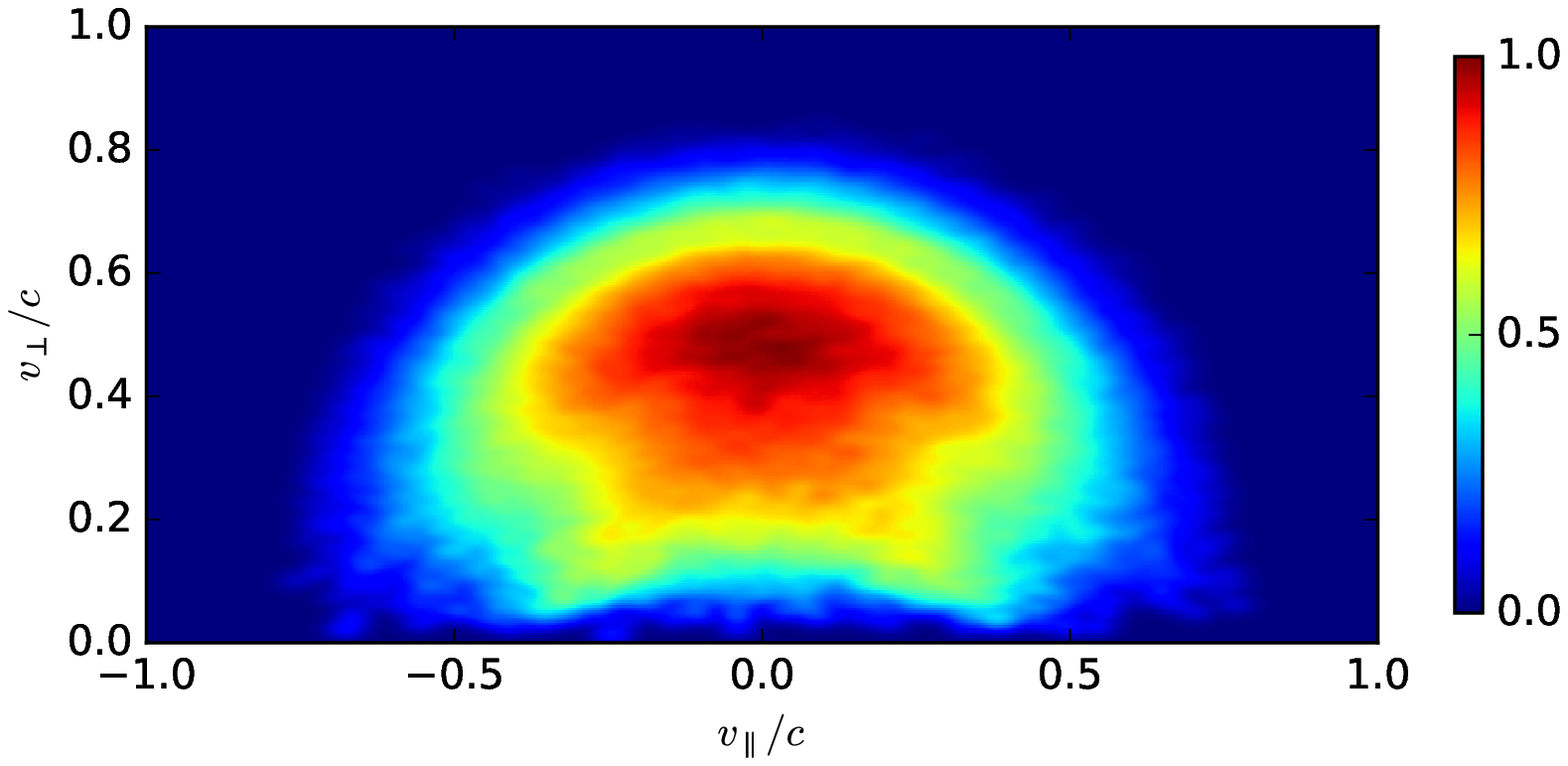}\\
\includegraphics[width=0.49\textwidth]{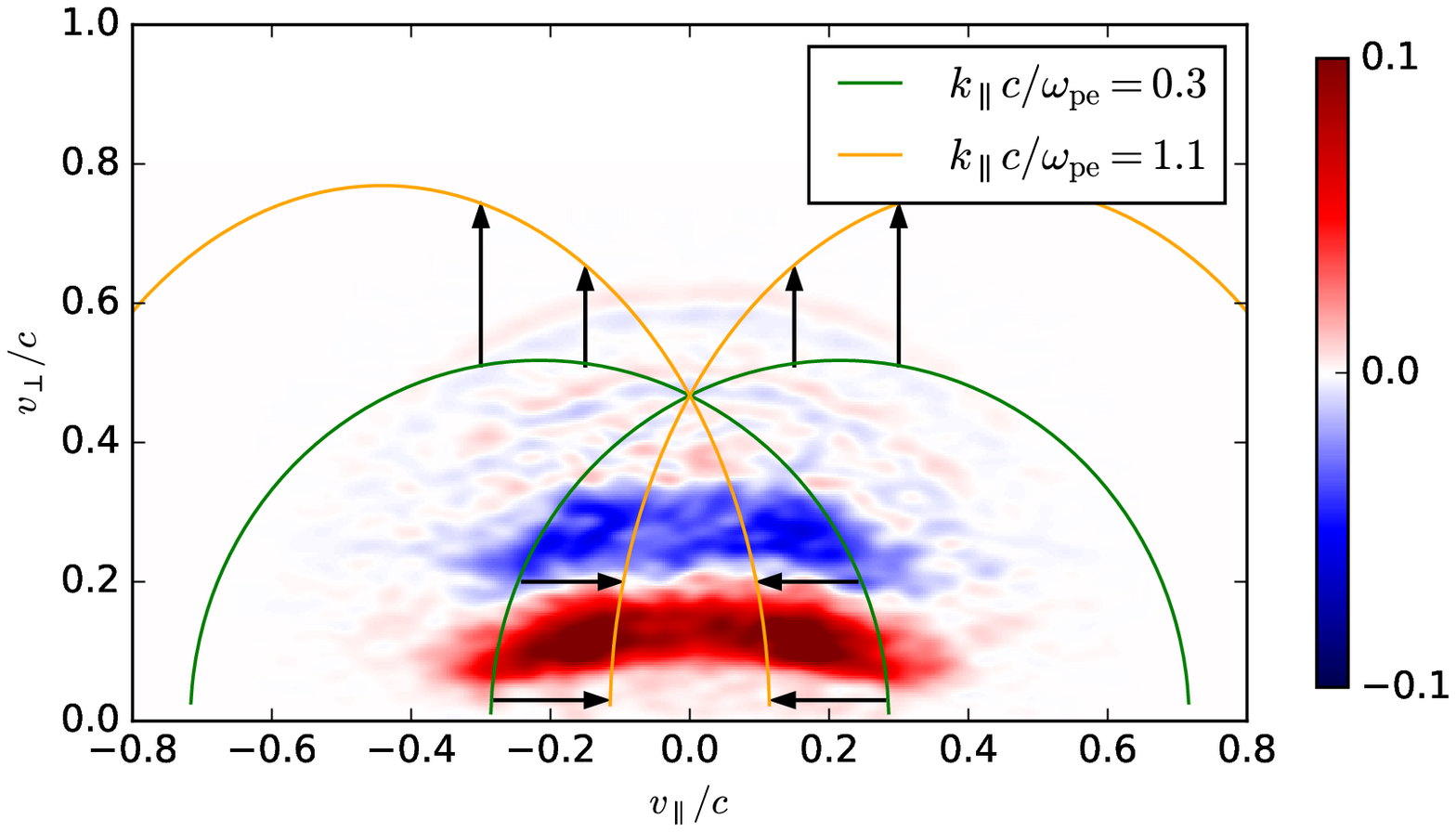}
\includegraphics[width=0.49\textwidth]{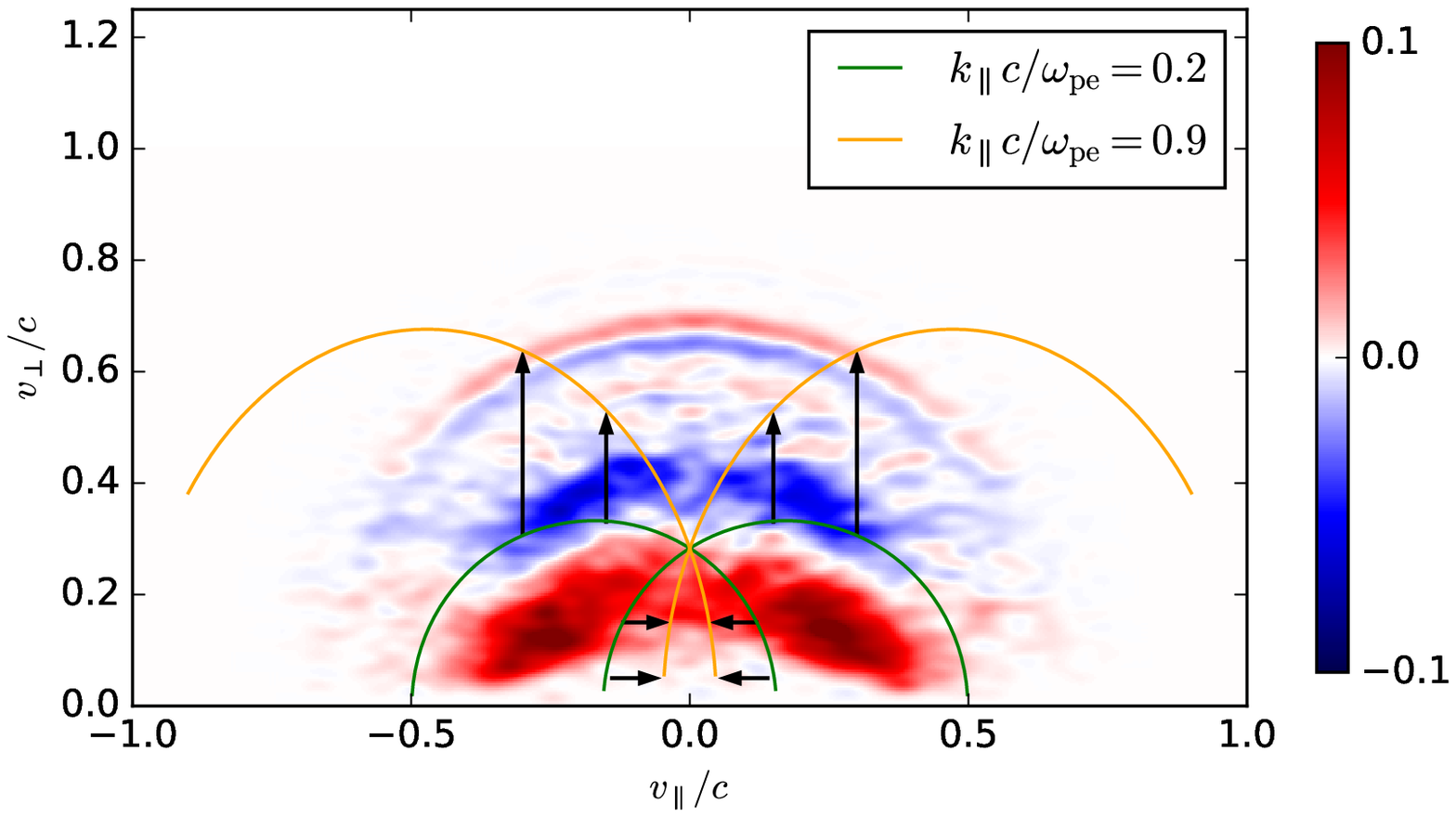}\\
\caption{Changes of the electron distribution functions of hot electrons during
the DPR instability. Left column: Model 2M with $s =6$. Right
column: Model 4M with resonance $s=5$. First row: The distribution at
the initial state. Second row: The distribution at
$\omega_\mathrm{pe}t = 1~000$. Distributions are normalized to their maximal
value. Third row: The difference of the distributions between the
initial state and at $\omega_\mathrm{pe}t = 1~000$. Red regions are with
enhanced densities and blue ones are with reduced densities. Elliptical
lines show the resonance ellipses for a given $k$-wave vector along magnetic
field. Black arrows show that the ellipses shift with increasing value of
$k_\parallel$. } \label{fig2}
\end{figure*}

To see the processes during the DPR instability in detail, we make a
comparison of the distributions in the initial state and at
$\omega_\mathrm{pe}t = 1~000$ for Models 2M and 4M (Table~\ref{tab1}; see
Figure \ref{fig2}). Comparing changes of the distribution in these models (left
and right columns in Figure \ref{fig2}), we can see that the instability with
different model parameters causes different changes of the distribution
function. The distribution function does not change only in one point of the
phase space, but in the subspace defined by resonance ellipses corresponding to
the range of $k_\parallel$. In Figure~\ref{fig2}, in the third row, we show
these resonance ellipses in the phase space, where the changes are dominant. We
also show how the resonance ellipses shift across the distribution function in
dependance on $k_\parallel$ (see the arrows). In both models, the loss-cones
of the distributions are step by step fulfilled by electrons (see the red
regions in the third row of Figure~\ref{fig2}) and thus the distributions
become closer to the Maxwellian distribution. These changes are due to an
increasing energy level of the upper-hybrid waves.

\begin{figure}[ht]
\centering
\includegraphics[width=0.45\textwidth]{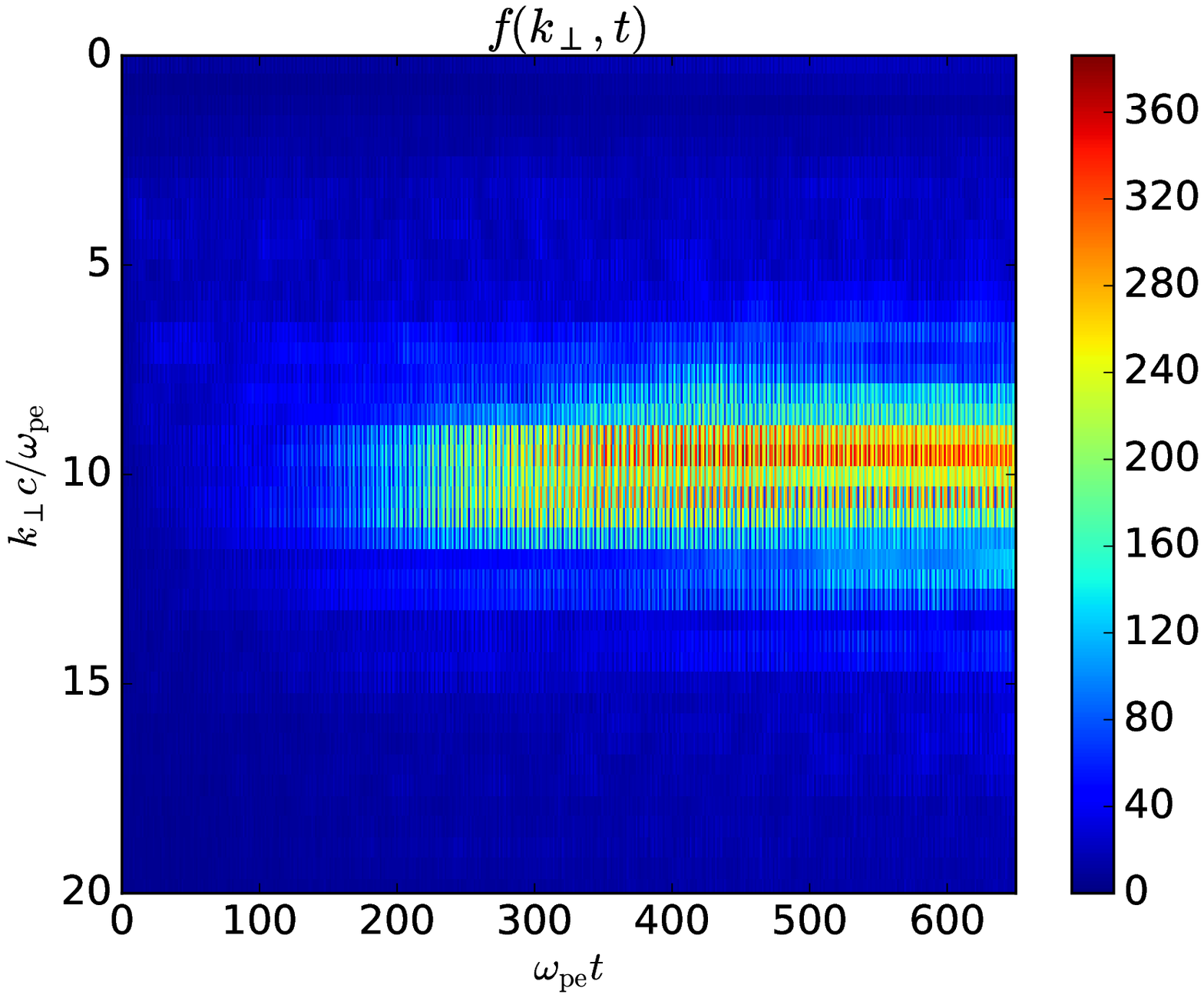}
\caption{Time evolution of energies in waves with the k-wave vectors in perpendicular direction to the magnetic field lines
in Model 2M.}
\label{fig11}
\end{figure}

We analyzed time evolution of energies in waves with the k-wave vectors
perpendicular to the magnetic field in Model 2M (Figure \ref{fig11}). As seen in
this figure, the interval of k-vectors is relatively narrow and remains
practically the same during the instability evolution. It shows that the energy
of the upper-hybrid waves is concentrated in the relatively narrow interval of
k-vectors. As will be shown in the following, it enables us to use the
specific-mode models.

\subsection{Comparison of specific-mode and multi-mode model saturation}

\begin{figure}[!ht]
\centering
\includegraphics[width=0.45\textwidth]{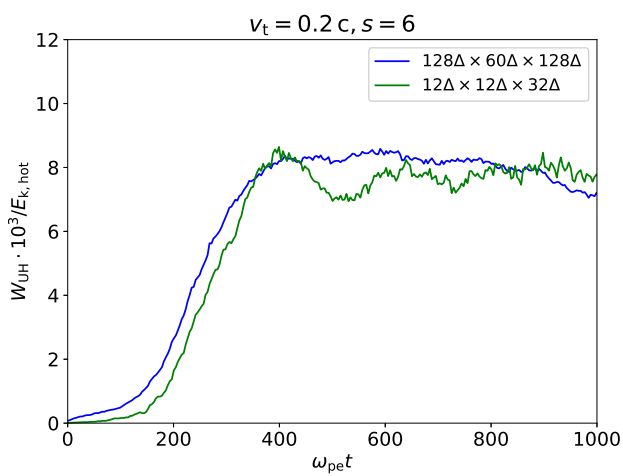}
\caption{Time evolution of the upper-hybrid wave energies for the multi-mode model~2M and
specific-mode model with the same model parameters.  }
\label{fig12}
\end{figure}

Because our main objective is to determine the growth rates and
saturation energies in a broad range of parameters (which is important for the
interpretation of observed zebras), we search for ways to accelerate
computations. We decided to use the specific-mode models that are much faster
than the multi-mode ones. To justify their use, we compare the growth
rate and saturation energies for all multi-mode models shown in
Table~\ref{tab1} with the specific-mode models with the same physical
parameters. An example of this comparison for Model~2M is shown in
Figure~\ref{fig12}. While in the multi-mode model we use the model size
$128\Delta \times 60\Delta \times 128\Delta$, in the specific-mode model in
this case we use the model size $12\Delta \times 12\Delta \times 32\Delta$.
This specific-mode model covers interval of $k_\perp c/\omega_{pe}$
above 5.23 (see Figure~\ref{fig11}). This means that this specific-mode model
covers the $k$-vector waves that are important for the DPR instability. This
explains why the results from the multi-mode and specific-mode models are very
similar (see the following).

As seen in Figure~\ref{fig12}, the growth rate and saturation energy in
both the models agree very well. The same result is found for all other models
according to Table~\ref{tab1}. This agreement is caused by a dominance of the
wave with the maximal growth rate (exponential increase) during an evolution of
the double plasma resonance instability. We utilize this agreement in the
following computations of the growth rates and saturation energies in a broad
range of the model parameters. The agreement between numerically and
analytically computed growth rates, shown in the following, also justifies the
use of the specific-mode models in our case.

\subsection{Effects of temperatures of hot and background plasma electrons on the growth rate}

\begin{figure}[!ht]
\centering
\includegraphics[width=0.49\textwidth]{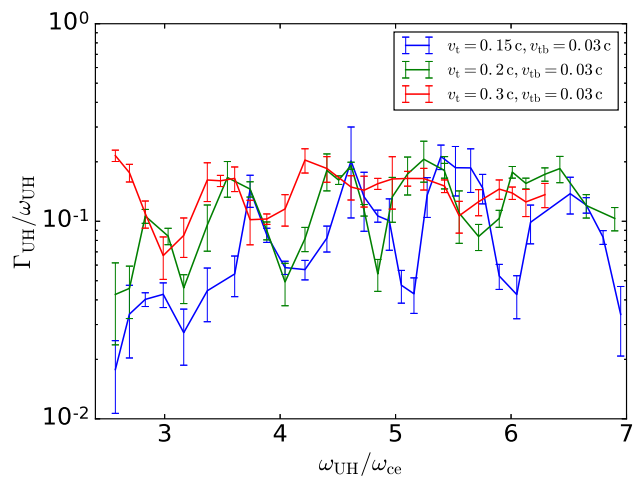}
\includegraphics[width=0.49\textwidth]{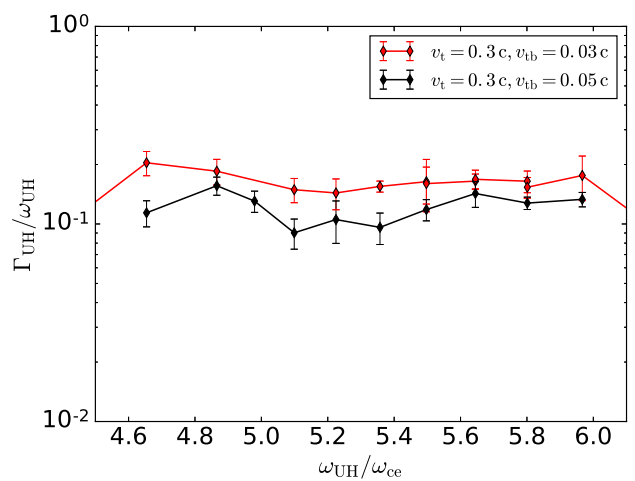}\\
\caption{Growth rates in dependence on $\omega_\mathrm{UH}/\omega_{ce}$ for Models 1S-4S.
Top: Plots for three different temperatures of hot electrons  $v_\mathrm{t}$.
Bottom: Plots for two different temperatures of background plasma electrons $v_\mathrm{tb}$. }
\label{fig3}
\end{figure}

In the following computations with the specific-mode models, we use the
model parameters as shown  in Table~\ref{tab2}. Figure \ref{fig3} presents
the effects of the temperature of hot and background electrons on the growth rate.
Every point in this figure is computed for at least three model sizes of the
specific-mode models. Error bars are estimated from their fit and the probability
that the same initial parameters give the same result. Models with higher
temperatures have lower errors because generated waves are more dominant over
the background noise.

\begin{table}[h!]
\caption{Parameters of the specific-mode models.} \label{tab2} \centering
\begin{tabular}{ccr}
\hline\hline
Model no. & $v_\mathrm{t}$ & $v_\mathrm{tb}$ \\
\hline
1S & 0.15~c & $0.03$~c \\
2S & 0.2~c & $0.03$~c \\
3S & 0.3~c & $0.03$~c \\
4S & 0.3~c & $0.05$~c \\
\hline
\end{tabular}
\end{table}

\begin{figure}[!ht]
\centering
\includegraphics[width=0.49\textwidth]{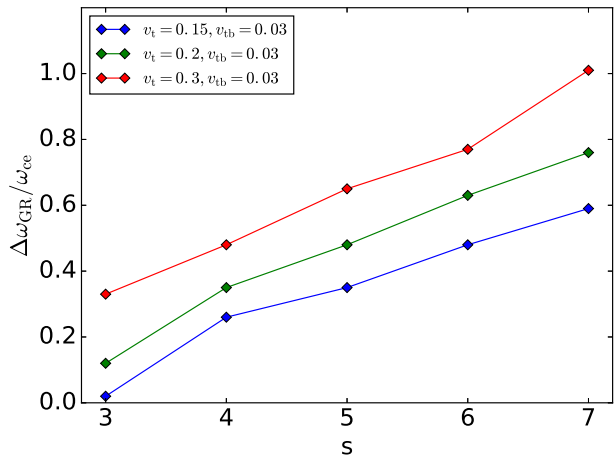}
\caption{Shifts of frequency of growth rate maxima from frequency given by gyro-harmonic number $s$.}
\label{fig4}
\end{figure}

Growth rates have the maxima that are shifted to frequencies lower than those
given by the gyro-harmonic number $s$ (Figure \ref{fig3}, upper part and
Figure~\ref{fig4}) and this shift increases with increasing temperature.
The ratio between maximal and minimal growth rates for each temperature is up
to one order and increases with the decreasing temperature of hot electrons. These
effects are in agreement with the analytical results
\citep{2017A&A...555A...1B}. However, contrary to the analytical results, the growth
rate for $v_\mathrm{t}$ = 0.3 is higher than that for lower temperatures. We
think that this difference is caused  by differences in the analytical and
numerical approach. While in the analytical approach we calculate the growth
rate only for one specific $\mathbf{k}$-vector wave, in numerical computations $\mathbf{k}$-vectors in some interval are always in operation.
The effect of the temperature of background plasma electrons is small (Figure
\ref{fig3}, bottom part), which agrees with the analytical theory
\citep{2017A&A...555A...1B}.
We also determined the shifts of maxima of growth rate (Figure~\ref{fig4}).
They increase with increasing $s$ and also depends on temperature of hot
electrons.

\subsection{Effect of the hot-cold plasma density ratio on the frequency of the growth rate maximum}

\begin{figure}[htb] \centering
\includegraphics[width=0.45\textwidth]{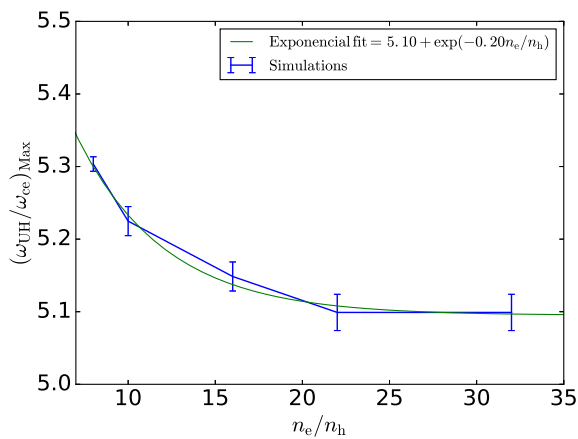}
\caption{Frequency of growth rate maximum
$(\omega_\mathrm{UH}/\omega_\mathrm{ce})_\mathrm{Max}$ as function of ratio
of background and hot electron densities $n_\mathrm{e}/n_\mathrm{h}$ for
parameters in Model~2S and $s=6$.} \label{fig1c}
\end{figure}

As shown in Figure~\ref{fig1c}, the frequency of the growth rate
maximum also depends on the ratio of the background and hot electron densities
$n_\mathrm{e}/n_\mathrm{h}$ .  This dependence is exponential as shown by its
fit and converges to $\omega_\mathrm{UH}/\omega_\mathrm{ce} = 5.1$ for high
values of the $n_\mathrm{e}/n_\mathrm{h}$ ratio.
Namely, low values of the $n_\mathrm{e}/ n_\mathrm{h}$ ratio shift the
resonance of the DPR instability and thus the frequency of the growth rate
maximum. In the analytical analysis it is supposed that the density
of cold electrons is much greater than that of hot electrons ($n_\mathrm{e} \gg
n_\mathrm{h}$).

In most of our simulations the density ratio
$n_\mathrm{h}/n_\mathrm{e}$ is 1:8 in order to keep a low numerical noise.
However, such dependencies enable us to extrapolate our results to much lower
density ratios, for example to 1:100 as usually considered in the zebra
interpretation (see the following).

\subsection{Comparison of the numerical and analytical growth rates}
\begin{figure}[ht]
\centering
\includegraphics[width=0.49\textwidth]{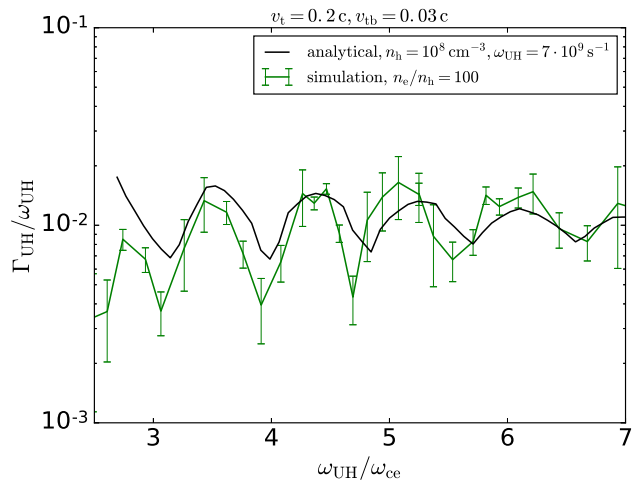}
\caption{Comparison of the simulated and analytically derived growth rates
for $v_\mathrm{t} = 0.2\,c$ and $v_\mathrm{tb} = 0.03\,c$.}
\label{fig10}
\end{figure}

We compared the simulated growth rates with the analytical ones presented in
the paper by \citet{2017A&A...555A...1B}. The comparison is shown on Figure
\ref{fig10}. It was made as follows. We chose the upper-hybrid frequency and
hot electron density as $\omega_\mathrm{UH} = 7\cdot10^{9}$~s$^{-1}$
($f_\mathrm{UH}=1.11$~GHz) and $n_\mathrm{h} = 10^8$~cm$^{-3}$. We changed the
growth rate plot expressed in dependence on $\omega_\mathrm{ce}$
\citep{2017A&A...555A...1B} to that in dependence on $\omega_\mathrm{UH}$.
Then, we transformed the simulated growth rates, computed for
$n_\mathrm{e}/n_\mathrm{h} = 8$,  to those with $n_\mathrm{e}/n_\mathrm{h} =
100$. The transformation was made in two steps. We shifted the simulated growth rate maxima
toward lower values of $\omega_\mathrm{UH}/\omega_\mathrm{ce}$ according to the
relation expressed in Figure~\ref{fig3}. Finally, the growth rates were
multiplied by the factor $8/100$ using the linear relation between the growth
rate and density of hot electrons \citep{2004SoPh..219..289Y}. As seen in
Figure \ref{fig10}, there is good agreement between the numerically and
analytically computed growth rates.

\subsection{Saturation energy levels of the upper-hybrid waves in a broad range of gyro-harmonic numbers}
In all models, we determined the saturation energy of the upper-hybrid waves.
The saturation energy is reached when the growth rate is balanced by nonlinear
effects.
The time when the saturation begins depends on the model parameters. In our cases, the saturation
time was in the time interval $\omega_\mathrm{pe}t_\mathrm{satur} = 300 -
1~200$. Mostly, the maximal growth rate leads to maximal saturation energy, but
this is not a rule. In some cases, the saturated energy is higher in the model
with lower growth rate. In other cases, the saturation energies have
recognizable maxima, while the growth rate is nearly constant in the broad
range of ratios of $\omega_\mathrm{UH}/\omega_\mathrm{ce}$.

\begin{figure}[h]
\centering
\includegraphics[width=0.49\textwidth]{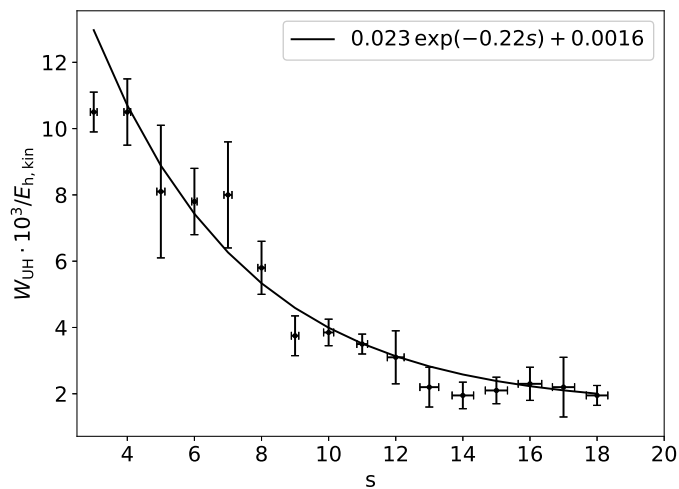}
\caption{Maximal saturation energies of the upper-hybrid waves
in dependence on the gyro-harmonic number $s$ for Model 2S with the exponential fit.}
\label{fig7}
\end{figure}

We computed one set of models changing the gyro-harmonic number $s$ from $s=3$
up to $s=18$ and keeping the same temperatures as in Model 2S. As shown in
Figure~\ref{fig7}, the saturation energy level in local maxima decreases with
increasing $s$ and converges to the value $W_\mathrm{UH} \approx
1.6\cdot10^{-3} E_\mathrm{h,kin}$ (see the exponential fit as presented in
Figure \ref{fig7}). This exponential function can be used for
an extrapolation of the saturation energies for even higher parameter $s$,
which is useful for zebras with many stripes.

Maximal values of the saturated energy levels of the upper-hybrid waves for low
values of $s$ are about one percent of the kinetic energy of hot electrons. The
saturated energy levels increase with an increase in the kinetic energy (temperature)
of hot electrons.

\begin{figure}[htb]
\centering
\includegraphics[width=0.49\textwidth]{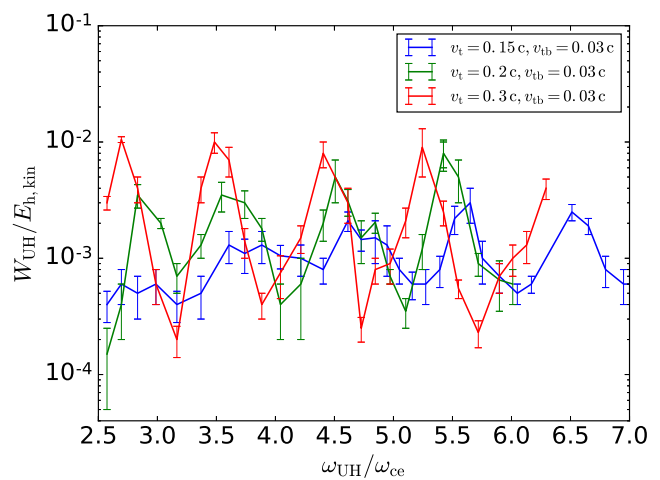}
\includegraphics[width=0.49\textwidth]{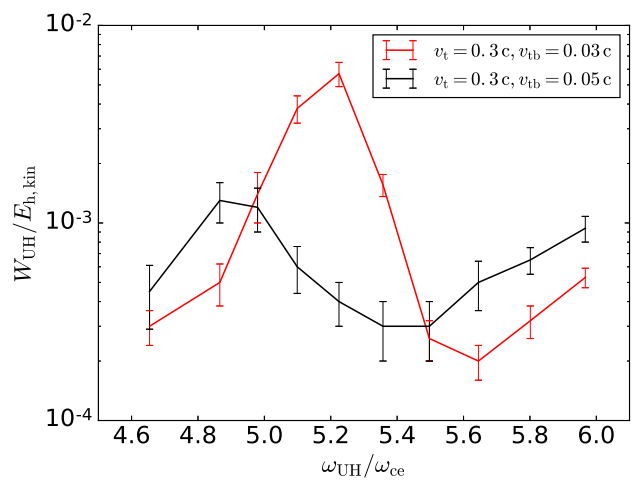}
\caption{Saturation energies as a function of $\omega_\mathrm{UH}/\omega_\mathrm{ce}$.
Top: Plots for three different temperatures of hot electrons  $v_\mathrm{t}$.
Saturation maxima are shifted to lower ratios $\omega_\mathrm{UH}/\omega_\mathrm{ce}$.
Bottom: Plots for two different temperatures of background plasma electrons $v_\mathrm{tb}$.}
\label{fig5}
\end{figure}

\begin{figure}[htb]
\centering
\includegraphics[width=0.49\textwidth]{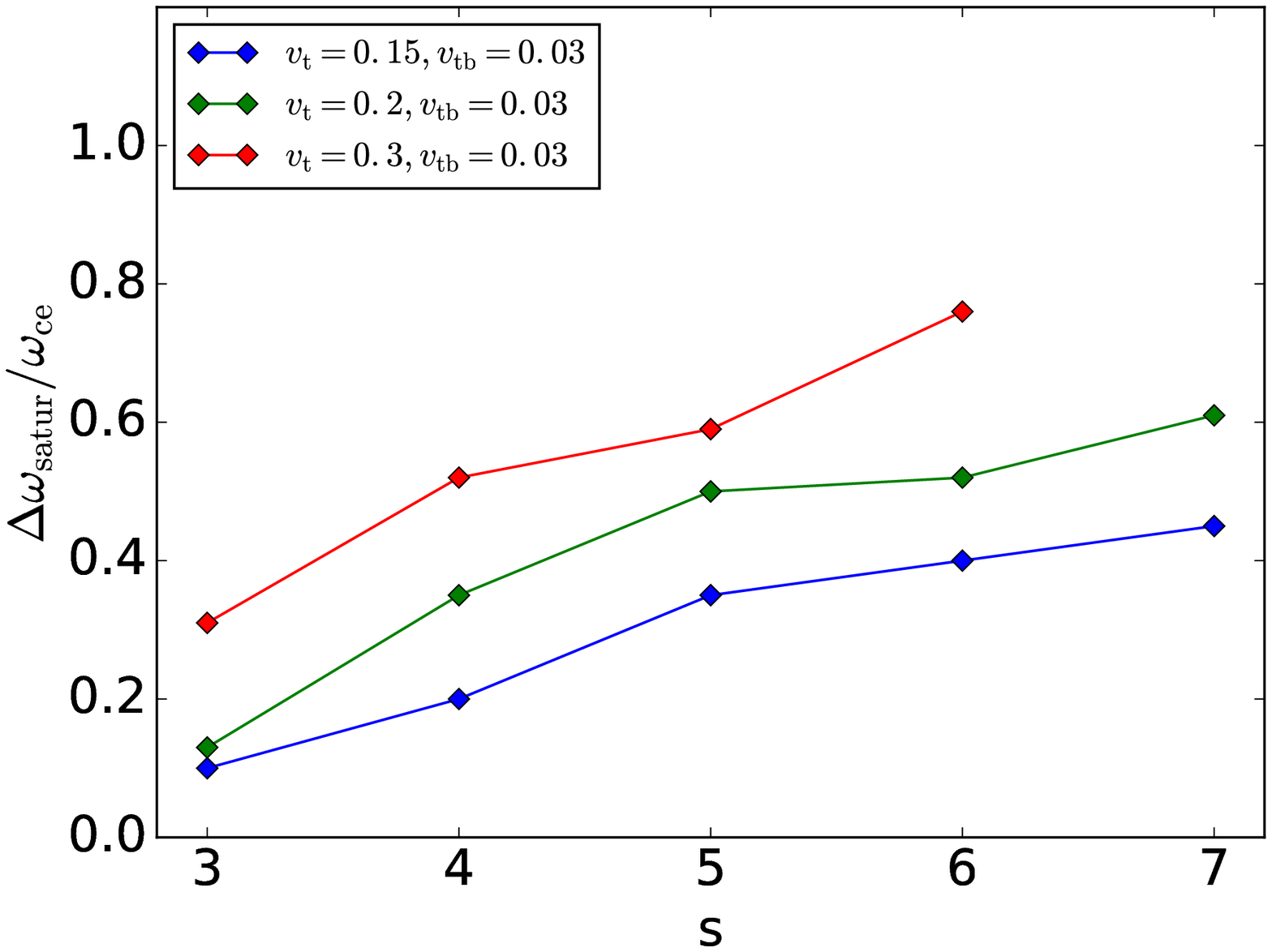}
\caption{Shifts of the frequency of the saturated energy maxima from the frequency given by the gyro-harmonic number $s$.}
\label{fig6}
\end{figure}

Similarly to the growth rates, the saturation energies  (Figure \ref{fig5}) have
the maxima and minima shifted to lower ratios of
$\omega_\mathrm{UH}/\omega_\mathrm{ce}$ than those given by the integer values of the
gyro-harmonic number $s$. Figure \ref{fig6} shows these shifts. The
frequency of the growth rate and saturation energy maximum is not generally the
same. The saturation energy maxima are usually less shifted to lower
frequencies than the growth rate maxima. However, contrary to the growth rates, the
contrast between the maxima and minima increases with the increasing
temperature of hot electrons (Figure \ref{fig5}, upper part). Similarly, in the
models that vary the temperature of background plasma electrons, the contrast
between the saturation energy maxima and minima is higher than that for the
growth rates (Figure \ref{fig5}, bottom part).

\section{Discussion and conclusions}

Using the 3-D PIC model in two versions (multi-mode and specific-mode
models) we computed not only the growth rates of the double plasma resonance
instability, but also saturation energies of the generated upper-hybrid waves.
We described details of the DPR instability by showing how the
distribution function of hot electrons changes during the DPR instability.
We found that the growth rate as well as the saturation energy are
proportional to the density of hot electrons.

Owing to many assumptions made in our previous analytical study, we
checked the analytical results using the present numerical models. We found a
very good agreement between the numerical and analytical results. This
justifies a use of the specific-mode models in this case. We confirmed that the
growth rate maxima are shifted to lower frequencies in comparison with those
given by the gyro-harmonic number $s$. This frequency shift increases with an
increase of the temperature of hot electrons, in agreement with the analytical
results. We confirmed that the contrast between maxima and minima of the growth
rate decreases with the increasing of the hot electron temperature. On the other
hand, the temperature of the background plasma has only a small effect on the
growth rates.

We found that the frequency of the growth rate maximum depends on the ratio of the
background and hot electron densities $n_\mathrm{e}/n_\mathrm{h}$. We used this
dependence
 to extrapolate our numerical results, made mostly
for the ratio $n_\mathrm{e}/n_\mathrm{h}$ = 8, to those with the ratio
$n_\mathrm{e}/n_\mathrm{h}$ = 100 considered usually in the analytical studies
of zebras. Using this dependence in detailed comparison of the analytical and
numerical growth rates, made in the interval of
$\omega_\mathrm{UH}/\omega_\mathrm{ce}$ = 3-7 and the parameters considered in
zebras, we found very good agreement.

We think that some small differences between the numerical and
analytical results are caused by differences in the two methods. In numerical
simulations, as in reality, the DPR instability works in some regions
of the $k$-vector space, not only with one $k$-vector as assumed in the
analytical theory. Furthermore, in some of the present simulations we found deviations
from the assumption made in the analytical approach ($k_\perp \gg
k_\parallel$).

We computed the saturation energies of the generated upper-hybrid
waves, which is beyond the possibilities of the analytical theory. We compared
the results of the multi-mode and specific-mode models considering the same
physical parameters. We found very good agreement between the results from
both types of model. This agreement is caused by a dominance of the wave with
the maximal growth rate during an evolution of the double plasma resonance
instability. Therefore we used the specific-mode models, which considerably
save computational time, for the computation of the saturation energies in a
broad range of the parameter $s$. A use of the specific-mode models is also
justified by a very good agreement between the growth rates computed
numerically and analytically. We found that the saturation energy decreases
with increasing $s$ and this decrease is exponential. This exponential
dependance of the saturation energy can be used for an extrapolation of the
saturation energies for the parameter $s$ even greater than 18, which is useful
for the interpretation of zebras with many zebra stripes
\citep{2015A&A...581A.115K,2015SoPh..290.2001Y}.

\begin{acknowledgements}
We acknowledge support from Grants 16-13277S and 17-16447S of the Grant Agency
of the Czech Republic. Computational resources were provided by the CESNET
LM2015042 and the CERIT Scientific Cloud LM2015085, provided under the
programme "Projects of Large Research, Development, and Innovations
Infrastructures".
\end{acknowledgements}

\bibliographystyle{aa}
\bibliography{DPR}

\end{document}